\documentclass[epj]{svjour}
\usepackage{epsfig}
\usepackage{color}
\newcommand{\ful}{\mbox{C$_{\mbox{\scriptsize{60}}}$}}
\newcommand{\vscf}{\mbox{$\delta V$}}
\newcommand{\eq}[1]{Eq.~\ref{#1}}
\begin{document}
\title{Coherence of Auger and inter-Coulombic decay processes in the photoionization of Ar@C$_{60}$ versus Kr@C$_{60}$}
\titlerunning{Auger-ICD coherence in Ar@C$_{60}$ vs Kr@C$_{60}$}
\author{Maia Magrakvelidze\inst{1}\thanks{Current address: Department of Physics, Kansas State University, Manhattan, Kansas 66502, USA} \and Ruma De\inst{1} \and Mohammad H. Javani\inst{2} \and Mohamed E. Madjet\inst{3} \and Steven T. Manson\inst{2} \and Himadri S. Chakraborty\inst{1} 
}                     
%
%
\institute{D.L. Hubbard Center for Innovation and Entrepreneurship, Department of Chemistry and Physics, Northwest Missouri State University, Maryville, Missouri 64468, USA, \email{himadri@nwmissouri.edu} \and Department Physics and Astronomy, Georgia State University, Atlanta, Georgia 30303, USA \and  Qatar Environment and Energy Research Institute, Hamad Bin Khalifa University, Qatar Foundation, P.O. Box 5825, Doha, Qatar}
\date{Received: date / Revised version: date}
%
\abstract{For the asymmetric spherical dimer of an endohedrally confined atom and a host fullerene, an innershell vacancy of either system can decay through the continuum of an outer electron hybridized between the systems. Such decays, viewed as coherent superpositions of the single-center Auger and two-center inter-Coulombic (ICD) amplitudes, are found to govern leading decay mechanisms in noble-gas endofullerenes, and are likely omnipresent in this class of nanomolecules. A comparison between resulting autoionizing resonances calculated in the photoionization of Ar@C$_{60}$ and Kr@C$_{60}$ exhibits details of the underlying processes. 
} 
\maketitle
\section{Introduction}
\label{sec1}
For a single-center system (generally an atom), the decay of an innershell electronic vacancy through an outershell continuum is the standard Auger process where the {\em intra}-Coulombic correlation enables local energy transfer from the de-excitation to the ionization process. For multi-centered systems, like molecules, dimers or polymers, a non-local energy transfer can dominate, namely, the decay of a hole at one center, inducing the emission of an electron from another - the {\em inter}-Coulombic decay (ICD) process \cite{cederbaum1997firstTh,marburger2003firstExp}. This process is stronger and cleaner if the bonding between monomers are weak. Over last several years, considerable theoretical \cite{averbukh2011revTh} and experimental \cite{hergenhahn2011revExp,jahnke2015rev} efforts have gone into ICD studies using rare gas dimers \cite{jahnke2004rareDimer}, rare gas clusters \cite{oehrwall2004rareClust}, surfaces \cite{grieves2011surface}, and water droplets \cite{jahnke2010water1,mucke2010water2}. Ultrafast ICDs of a dicationic monomer in a cluster to produce a cluster tricataion \cite{santra2003triCataion} or multiply excited homoatomic cluster \cite{kuleff2010multiExcited} were predicted. Relatively recently, ICD following the resonant Auger decay is identified in Ar dimers using momentum resolved electron-ion-ion coincidence spectroscopy \cite{okeeffe2013arDimer,kimura2013arDimer}. Furthermore, experiments are also possible nowadays to probe the temporal aspects of ICD mechanism in matters \cite{fruehling2015time-icd}. In fact, time domain measurements of ICD in He \cite{trinter2013time1} and Ne \cite{schnorr2013time2} dimers have recently been achieved. In the context of medical applications, specifically radio-oncology, the low-energy ICD process was discussed \cite{gokhberg2014MedApp,trinter2014MedApp}.

In this paper, we are interested in the resonant ICD (RICD) process where the initial vacancy is induced by the absorption of a photon causing an innershell photoexcitation. Contemporary research has addressed various small clusters and dimers to unravel effects of photon-stimulated RICD. A prediction of strong RICD activities following Ne $2s\rightarrow np$ excitations in MgNe clusters was made about a decade ago \cite{gokhberg2006ricdTh}. Experimentally, evidence for RICD was seen in the photoelectron spectroscopy of Ne clusters for $2s\rightarrow np$ excitations \cite{barth2005ricdExp1}, and also in the double photoionization of Ne dimers by monitoring the creation of energetic Ne$^+$ \cite{aoto2006ricdExp2}. Strong enhancement of the HeNe$^+$ yield, as He resonantly couples with the radiation, was recently detected \cite{trinter13HeNe}, confirming an earlier prediction \cite{najjari10TwoCenter}.

Atoms endohedrally confined in fullerene molecules, endofullerenes, being near-spherical, atom-cluster dimers of loose Van-der-Walls type bonding have attracted significant attention as natural laboratories for ICD research. These materials are stable in the room temperature with inexpensive sustenance cost and their synthesis techniques are also rapidly improving. The earliest attempt to predict ICD in endofullerenes was made by calculating ICD rates for Ne@$\ful$ \cite{averbukh2006endo-icdTh}. This was followed by some studies of Coulomb-mediated energy transfer from atom to fullerene that broaden the Auger lines \cite{korol2011,amusia2006}. Systems supporting regular RICD can be visualized as antenna-receiver pairs at the molecular scale \cite{trinter13HeNe} where the antenna couples to the incoming radiation and transfers energy to the receiver to perform an act of emission. Very recently, a different class of resonances decaying into atom-fullerene hybrid final state vacancies for the photoionization of Ar@$\ful$ have been predicted; this arises from the competition of the intra-Coulombic Auger channel with an intrinsically connected ICD channel \cite{javani2014-rhaicd}. The calculated features were found to be remarkably stronger than both regular ICD and Auger resonances.  Obviously, these processes can accentuate the system efficiency by enabling the antenna to also contribute to the emission resonantly with the receiver through the coherence. Therefore, given that this effect may have utilization in nanoscale antenna technology \cite{novotny2012NanoAntenna} besides its established basic-science context, it is of great interest to investigate if such coherence phenomena are a common place energy-transfer mechanism in the spectroscopy of endofullerenes. 

To this purpose, we extend our calculations to a number of noble gas endofullerenes and find strong abundance of such coherence in the spectral landscape of these systems. In this paper, we compare between the results of Ar@$\ful$ and Kr@$\ful$ to uncover details of the process; particularly, the dependence of the spectral features on the choice of the encapsulated atom. Section 2 carries two subsections, providing a short account of the theoretical method, predicting atom-fullerene hybridization, and an interchannel coupling based description of the Auger-ICD coherence. Section 3 presents the final numerical results of the resonances with discussions. We conclude the paper in Section 4.

\section{Theoretical Details}
\subsection{The methodology}
\label{sec2-1}
Kohn-Sham density functional theory is used to describe the ground state electronic structure of the compounds using same methodology employed earlier \cite{madjet2010xeFull}. The $\ful$ molecule is modeled by smearing sixty C$^{4+}$ ions into a spherical jellium shell, fixed in space, with an experimentally known $\ful$ mean radius 3.5 \AA and thickness $\Delta$, augmented by a constant potential $V_0$. The nucleus of the confined atom is placed at the center of the sphere. The Kohn-Sham equations for the system of a total of $240+N$ electrons ($N=18$ for Ar, $N=36$ for Kr and 240 delocalized electrons from $\ful$) are then solved to obtain the electronic ground state properties in the local density approximation (LDA). The gradient-corrected Leeuwen and Baerends exchange-correlation functional [LB94]~\cite{van1994exchange} is used for the accurate asymptotic behavior of the ground state radial potential
\begin{equation}\label{lda-pot}
V_{\scriptsize \mbox{LDA}}(\mathbf{r}) = -\frac{z}{r} + \int d\mathbf{r}'\frac{\rho(\mathbf{r}')}{|\mathbf{r}-\mathbf{r}'|} + V_{\scriptsize \mbox{XC}}[\rho(\mathbf{r})],
\end{equation} 
which is solved self-consistently in a mean-field framework. The parameters $V_0$ and $\Delta$ are determined by requiring both charge neutrality and obtaining the experimental value, 7.54 eV, of the first $\ful$ ionization potential. This procedure yields a value of $\Delta$ of 1.3\AA, in agreement with the value inferred from experiment \cite{ruedel2002oscExp}.

Significant ground state hybridization of atomic valence orbitals $np$ ($n=3,4$ respectively for Ar and Kr) with $\ful$ $3p$ is found, resulting in (X$np$$\pm\ful$$3p$) levels from the symmetric and antisymmetric mixing similar to the bonding and antibonding states in molecules or dimers:
\begin{equation}\label{bound-hyb}
\mbox{X}np\!\pm\!\ful 3p = |\phi_\pm\rangle = \sqrt{\alpha}|\phi_{np \scriptsize{\mbox{X}}}\rangle \!\pm\! \sqrt{1-\alpha}|\phi_{3p \scriptsize{\mbox{C}_{60}}}\rangle,
\end{equation} 
where X denotes Ar and Kr. The radial wavefunctions corresponding to these levels and their binding energies are shown in Figure 1(a). Note that in the Ar case the anti-symmetric combination induces one fewer node and is more strongly bound compared to the symmetric, while the opposite is true for Kr@$\ful$. Such atom-fullerene hybridization was predicted earlier \cite{chakraborty2009xe-hybrid} and detected in a photoemission experiment on multilayers of Ar@$\ful$ \cite{morscher2010strong}. In fact, the hybridization gap of 1.52 eV between (Ar+$\ful$) and (Ar$-\ful$) in our calculation is in good agreement with the measured value of 1.6$\pm$0.2 eV \cite{morscher2010strong}. Outside these hybrids, we use the symbol $n\ell$@ to denote the levels of the confined atom and @$n\ell$ to represent the levels of the doped $\ful$.
\begin{figure}
\centerline{\psfig{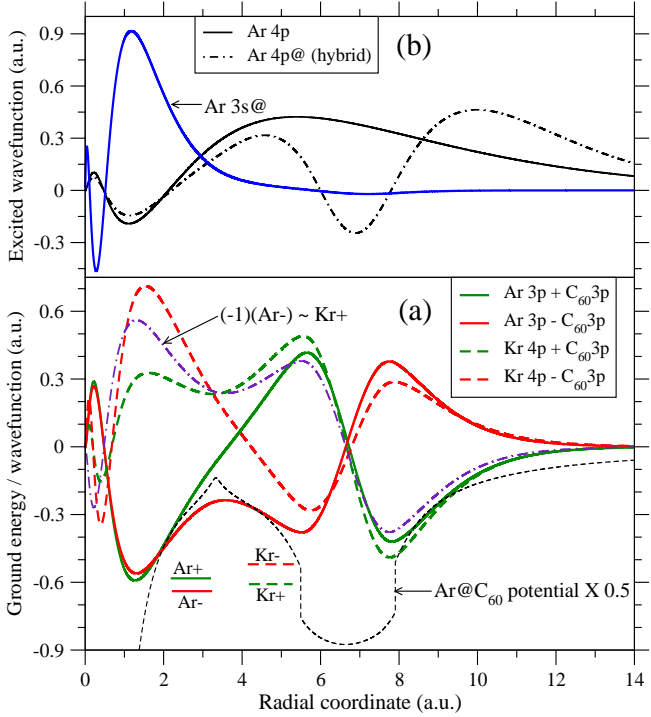}}
\caption{(Color online) (a) Ground state radial wavefunctions and binding energies of Ar@$\ful$ and Kr@$\ful$ hybrid levels; X$\pm$ are used for short-hand notations for these levels where X is Ar or Kr. That the inverted Ar- is quite similar to Kr+ is noted. The radial potential of Ar@$\ful$ is also shown.  (b) The excited $4p$ wavefunctions of free and confined Ar, and the inner $3s$ wavefunction of confined Ar are plotted.}
\label{fig1}       
\end{figure}

A time-dependent LDA (TDLDA) approach \cite{madjet-jpb-08} is used to calculate the dynamical response of the compound to the external dipole field $z$. In this method, the photoionization cross section corresponding to a bound-to-continuum dipole transition $n\ell\rightarrow k\ell^\prime$ is
\begin{equation}\label{cross-pi}
\sigma_{n\ell\rightarrow k\ell'} \sim |\langle k\ell'|z+\vscf|n\ell\rangle|^2,
\end{equation}
where the matrix element ${\cal M} = {\cal D} + \langle\vscf\rangle$, with ${\cal D}$ being the independent-particle LDA matrix element. 
Here $\delta V$ represents the complex induced potentials that account for electron correlations. In the TDLDA, $z + \delta V$ are proportional to the induced frequency-dependent changes in the electron density~\cite{madjet-jpb-08}.
This change is 
\begin{equation}\label{ind-dens}
\delta \rho (\mathbf{r}^{\prime}; \omega) = \int \chi (\mathbf{r}, \mathbf{r}^{\prime}; \omega)
z  d\mathbf{r},
\end{equation}
where the full susceptibility $\chi$ builds the dynamical correlation from the LDA susceptibilities, 
\begin{eqnarray}\label{suscep}
\chi^{0} (\mathbf{r},\mathbf{r}^{\prime };\omega) &=&\sum_{nl}^{occ}\phi _{nl}^{*}
(\mathbf{r})\phi _{nl}(\mathbf{r}^{\prime })\ G(\mathbf{r},\mathbf{r}^{\prime };\epsilon
_{nl}+\omega)  \nonumber \\
&+&\sum_{nl}^{occ}\phi _{nl}(\mathbf{r})\phi _{nl}^{*}(\mathbf{r}^{\prime })\ G^*
(\mathbf{r},\mathbf{r}^{\prime };\epsilon _{nl}-\omega)  
\end{eqnarray}
$via$   the matrix equation $\chi = \chi^0[1-(\partial V/\partial \rho)\chi^0]^{-1}$ involving the variation of the ground-state potential $V$ with respect to the ground-state density $\rho$. The radial components of the full Green's functions in \eq{suscep} are constructed with the regular ($f_L$) and irregular ($g_L$) solutions of the homogeneous radial equation 
\begin{equation}\label{radial-eq}
\left( \frac{1}{r^2} \frac{\partial}{\partial r} r^2 \frac{\partial}{\partial r} 
     - \frac{L(L+1)}{r^2} - V_{\mbox{\scriptsize{LDA}}} 
     + E \right) f_L(g_L) (r;E) = 0
\end{equation}
as
\begin{equation}\label{green}
G_{L}(r,r^{\prime };E)=\frac{2f_{L}(r_{<};E)h_{L}(r_{>};E)}{W [f_{L},h_{L}]}  
\end{equation}
where $W$ represents the Wronskian and $h_{L} = g_{L} + i\; f_{L}$. Obviously, TDLDA thus includes the dynamical correlation by improving upon the mean-field LDA basis.

\subsection{The description of Auger-ICD coherence}
\label{sec2-2}
The TDLDA matrix elements ${\cal M}$ for the dipole photoionization of (X$\pm$$\ful$) levels, in the interchannel coupling framework introduced by Fano \cite{fano1961}, can be written as \cite{javani12alkaline-earth},
\begin{eqnarray}\label{gen-mat-element}
{\cal M}_\pm (E) &=& {\cal D}_\pm (E) + {M}^{c-c}_\pm (E)+{M}^{d-c}_\pm (E),
\end{eqnarray}
where the single electron (LDA) matrix element ${\cal D}_\pm (E) = \langle ks(d)|z|\phi_\pm\rangle$; ${M}^{c-c}$ and ${M}^{d-c}$ are respectively corrections from continuum-continuum and discrete-continuum channel couplings, accounting for $\langle\vscf\rangle$ in \eq{cross-pi}. ${M}^{c-c}$ constitutes rather smooth many-body contribution to nonresonant cross section, while the resonance structures originate from ${M}^{d-c}$. Following \cite{fano1961},
\begin{eqnarray}\label{dc-mat-element}
 {M}^{d-c}_\pm &=& \displaystyle\sum_{n\ell} \sum_{\eta\lambda} \frac{\langle\psi_{n\ell\rightarrow\eta\lambda}|\frac{1}{|{\bf r}_{\pm}-{\bf r}_{n\ell}|}
|\psi_{\pm}(E)\rangle}{E-E_{n\ell\rightarrow\eta\lambda}} {\cal D}_{n\ell\rightarrow\eta\lambda},
\end{eqnarray}
in which the $|\psi\rangle$ refer to interacting discrete (inner) $n\ell\rightarrow\eta\lambda$ and continuum (outer) X$\pm$$\ful$$3p\rightarrow ks(d)$ {\em channel} wavefunctions; $E_{n\ell\rightarrow\eta\lambda}$ and ${\cal D}_{n\ell\rightarrow\eta\lambda}$ are LDA bound-to-bound excitation energies and matrix elements, respectively. The two-body interchannel coupling matrix elements (ICME) of the Coulomb interaction in \eq{dc-mat-element} are the conduits of the energy transfer process between channels. The excited states of the system are found hybridized [see Fig.\,1(b)], implying that innershell electrons from pure levels are excited to the hybrid levels. But we do not expect significant differences in ${\cal D}_{3s\rightarrow\eta p}$ between free and confined Ar. This is because, even though hybrid excited state wavefunctions have induced structures in the vicinity of the $\ful$ shell, the Ar $3s$@ wavefunction being un-mixed and localized on Ar [Fig.\,1(b)] means that its overlap with the excited state wavefunctions is largely unaffected by the hybridization. An identical reason also ensures that the doped $\ful$'s innershell excitation LDA matrix elements are also essentially unchanged.
\begin{figure}
\centerline{\psfig{figure=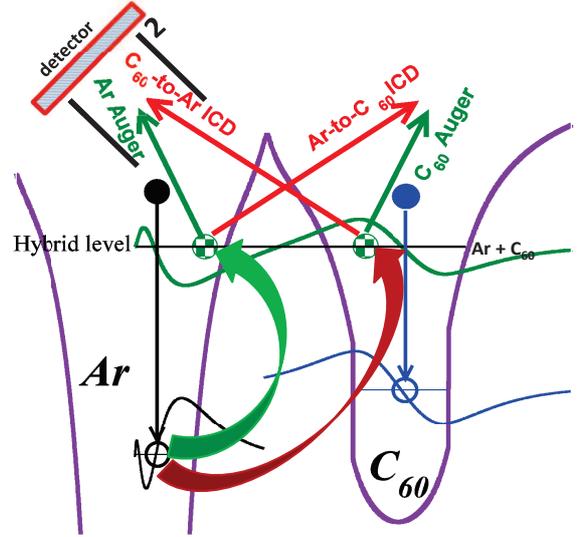,height=7.5cm,width=8.5cm,angle=0}}
\caption{(Color online) Schematic of coherent mixing of one-center Auger decay amplitudes (green) of core vacancies with corresponding cross-center ICD amplitudes (red) in the spectra of the Ar-$\ful$ hybrid electron. See text for a fuller description.}
\label{fig2}       
\end{figure}

Following \eq{bound-hyb}, the hybridization of the continuum channels in \eq{dc-mat-element} assumes the form
\begin{equation}\label{channel-hyb}
|\psi_\pm\rangle = \sqrt{\alpha}|\psi_{np@ \scriptsize{\mbox{X}}}\rangle \pm \sqrt{1-\alpha}|\psi_{@3p \scriptsize{\mbox{C}_{60}}}\rangle,
\end{equation}
where $\psi_{np@ \scriptsize{\mbox{X}}}$ and $\psi_{@3p \scriptsize{\mbox{C}_{60}}}$ are the wavefunctions of the channels arising, respectively, from the valence $np$ level of the atom and the $3p$ level of $\ful$. 
In Eqs.\,(\ref{channel-hyb}) we used @ to indicate the continuum waves in confined Ar and doped $\ful$. Using Eq.\ (\ref{channel-hyb}) in (\ref{dc-mat-element}), and recognizing that the overlap between a pure Ar and a pure $\ful$ bound state is negligible, we separate the atomic and fullerene regions of integration as
\begin{eqnarray}\label{dc-mat-element2}
{M}^{d-c}_\pm (E) &=& \displaystyle\sum_{n\ell} \sum_{\eta\lambda}\left[\sqrt{\alpha}\frac{\langle\psi_{n\ell\rightarrow\eta\lambda}|\frac{1}{|{\bf r}_{\pm}-{\bf r}_{n\ell}|}
|\psi_{np@ \scriptsize{\mbox{X}}}(E)\rangle}{E-E_{n\ell\rightarrow\eta\lambda}}\right. \nonumber \\
               \!\!&\!\!\pm\!&\!\left. \sqrt{1-\alpha} \frac{\langle\psi_{n\ell\rightarrow\eta\lambda}|\frac{1}{|{\bf r}_{\pm}-{\bf r}_{n\ell}|}
|\psi_{@3p \scriptsize{\mbox{C}_{60}}}(E)\rangle}{E-E_{n\ell\rightarrow\eta\lambda}} \right]\!\!{\cal D}_{n\ell\rightarrow\eta\lambda}. \nonumber \\
\end{eqnarray}

\eq{dc-mat-element2} can be schematically understood using Figure 2 with the example of Ar@$\ful$. If $n\ell\rightarrow\eta\lambda$ produces an Ar innershell hole, then the de-excitation process (black arrow) releases energy that can transfer to a hybrid level as if into two branches (thick curved arrows): (i) The first (green) is a local transfer that liberates the atomic part of the hybrid electron denoted by the first term in Eq.\,(\ref{dc-mat-element2}). This represents the ordinary Auger decay in Ar.  (ii) The second (red) is a non-local Ar-to-$\ful$ ICD energy transfer that knocks off the $\ful$ part of the same electron represented by the second term in Eq.\,(\ref{dc-mat-element2}). The {\em partial} electrons are denoted by the checkered spheres in Fig.\,2. For the photoionization cross section, which involves the modulus squared of the matrix element, these Auger and ICD components of the amplitude coherently mix to induce the resonance, resulting in a shared (hybridized) outershell vacancy. Likewise, for the de-excitation (blue arrow) of an original $\ful$ innershell hole, the first and second terms in Eq.\,(\ref{dc-mat-element2}) indicate the coherence between, respectively, a $\ful$-to-Ar ICD and a $\ful$ Auger process. Hence, these decay pathways can be termed a resonant hybrid Auger-inter-Coulombic decay (RHA-ICD).

\section{Results and discussion}
\label{sec3}
Figure 3(a) presents some selected Auger resonances for free atoms and the empty $\ful$ molecule. For each atom, the two lowest $ns\rightarrow np$ resonances in the valence cross sections are denoted by A and B. These are Ar $3s\rightarrow 4p,5p$ and Kr $4s\rightarrow 5p,6p$. These resonances are characteristically near-symmetric, window-type. Note that for each atom, the resonance width becomes significantly smaller with higher final states. This behavior relates to the property of the excited orbitals, bulk of whose amplitudes progressively move farther in the radial coordinate to weaken the overlaps in ICME similar in \eq{dc-mat-element}, because it involves a two-electron Coulomb operator. Furthermore, this property of decreasing width with increasing excitation follows directly from quantum defect theory \cite{Seaton83} which shows that the widths drop off as 1/n$^{*3}$ with increasing n$^*$, the effective principal quantum number.  Note further that five $\ful$Auger resonances are also identified in Fig.\,3(a) in the cross section of $\ful$ $3p$ (that participates in the hybridization [\eq{bound-hyb}] in endofullerenes) which are of diverse shapes with 1-4 being strongly asymmetric and a near-symmetric 5. As a rule, $\ful$ Auger resonances are narrower than atomic resonances \cite{madjet-jpb-08} which follows directly from the delocalized behavior of $\ful$ electrons since their being diffused in radial space produces smaller rates $via$ the ICME.   
\begin{figure*}
\centerline{\psfig{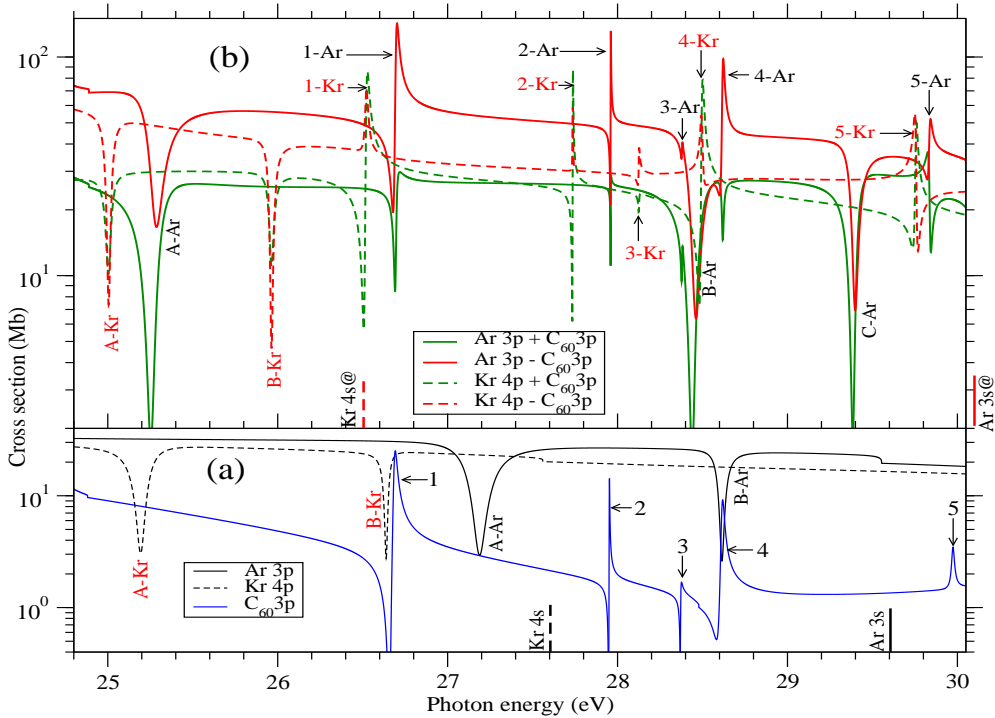}}
\caption{(Color online) (a) Photoionization cross sections of free Ar $3p$ and Kr $4p$ featuring two lowest Auger resonances for Ar $3s\rightarrow 4p,5p$ (A-Ar, B-Ar) and Kr $4s\rightarrow 5p,6p$ (A-Kr, B-Kr) innershell vacancies. Five Auger resonances, labeled 1-5, in $\ful$ $3p$ cross section are also shown. (b) The RHA-ICD counterparts of the resonances of panel (a) in the cross sections of hybrid electrons. Color-coded label tags (black for Ar and red for Kr) are used to guide the eye in identifying these resonances. Free and confined Ar $3s$ and Kr $4s$ ionization thresholds, to which the excitation series converge, are indicated in the respective panels.} 
\label{fig3}       
\end{figure*}

Figure 3(b) displays the photoionization cross sections, over the same energy range as Fig.\,3(a), of the endofullerene hybrid levels (X$\pm$$\ful$). The labeled resonances in this panel are the RHA-ICD ``avatars'' of the free-system Auger structures in Fig.\,3(a), and they occur almost at the same energies for each hybrid. This means that features A, B, and C (C only for Ar) in Fig.\,3(b) are resonances that emerge from the decay of Ar $3s@\rightarrow 4p@,5p@,6p@$ and Kr $4s@\rightarrow 5p@,6p@$ excitations through the continua of these hybrid levels. Remarkably, they are significantly stronger, particularly for anti-symmetric (X-$\ful$), than their Auger counterparts seen in Fig.\,3(a). In addition, another dramatic effect is evident: The resonances 1-5 in Fig.\,3(b), decaying through the hybrid continua, grow to an order of magnitude larger than the Auger resonances in empty $\ful$ [Fig.\,3(a)]. In essence, Ar and $\ful$ innershell vacancies decay significantly more powerfully through the photoionization continua of the X$\pm$$\ful$ hybrid levels than they do through the continua of pure levels. To understand why this happens, note that both the terms in Eq.\,(\ref{dc-mat-element2}) are large, owing to the substantial overlaps between innershell bound states and (X$\pm$$\ful$)$3p$ channel wavefunctions. But there is more. The resonances in the matrix element ${M}^{d-c}_\pm$ also coherently interfere with the nonresonant part ${\cal D}_\pm + {M}^{c-c}_\pm$, [Eq.\,(\ref{gen-mat-element})] which is generally stronger for hybrid levels \cite{maser2012zn-hybrid}. This interference, consequently, enhances RHA-ICD resonances compared to their Auger counterparts in free A or empty $\ful$ channels, as seen in Fig.\,3(b).

Resonances A and B [Fig.\,3(b)] of the confined Kr move lower in energy than the corresponding resonances [Fig.\,3(a)] for the free Kr, and this shift is greater for resonance B. For Ar too, such an energy red-shift is noted in moving from free to confined, but, in contrast to Kr, the shift in A in Ar is substantially larger. In order to understand this behavior, note first in Fig.\,(3) that the effect of confinement blue-shifts the inner Ar $3s$@ which is opposite to Kr $4s$@ that red-shifts under confinement. This can be explained as follows: In the compound system, the atom and $\ful$ exert mutual perturbations on each other. As a result, the general shift of energy levels from their ``unperturbed'' values is a function of two effects: (i) the addition of two attractive potentials should tend to make the levels more bound and (ii) the repulsion between the atomic and $\ful$ electron-groups in the self-consistent mean field induces just the opposite effect. It turns out that in our LDA ground state calculations, the former wins for Ar@$\ful$ but the later for Kr@$\ful$, since Kr adds a significantly larger number of electrons to the combined system. This enables Ar $3s$@ (-30.1 eV) and Kr $4s$@ (-26.5 eV) to become, respectively, more and less bound than their free results (-29.6 eV and -27.6 eV, respectively). In fact, this effect also shifts the entire excitation spectra in a similar fashion that can be seen in Table 1 where we determine the excited state energies from the resonance positions in Fig.\,(3) both for free and confined atoms. Note that free Ar and Kr excited state energies are essentially equal since their p-wave quantum defects differ by almost exactly 1 \cite{manson69,kennedy72,theodosiou86}, so it is only the confinement that somewhat complicates the results in Table 1. We should also keep in mind that the excited state energies for the compound systems are affected also by the orbital hybridization, resisting any simple systematics in the shift and, thus, in the positions of resonances A and B in Fig.\,3(b). Furthermore, Figs.\,3(a) versus (b) indicates that the positions of 1-5 RHA-ICD resonances for Ar@$\ful$ practically reproduce the positions of corresponding Auger lines, while those for Kr@$\ful$ are systematically red-shifted. Obviously, this is also due to the significantly larger electron-repulsion effects resulting in the lowering of ground-state level-energies in $\ful$ when the central atom is Kr, as discussed above.
%
\begin{table}
\caption{Energies (in eV) of free and confined excited states arising from ground Ar $3s$ and Kr $4s$ levels calculated from the positions of A and B in Fig.\,(3).}
\label{tab2}       
\begin{tabular}{l|llll}
\hline\noalign{\smallskip}
    & Ar & Ar@ & Kr & Kr@  \\
\noalign{\smallskip}\hline\noalign{\smallskip}
A & -2.4 (4p) & -4.8 & -2.4 (5p) & -1.5 \\
B & -1.0 (5p) & -1.7 & -1.0 (6p) & -0.5 \\
\noalign{\smallskip}\hline
\end{tabular}
\end{table}

It is also evident in Fig.\,3 that the RHA-ICD resonances A-C roughly retain the symmetric window shapes of their free atom Auger counterparts. This is because their atomic Auger components are playing the dominant role in the coherence. But the effect of the ICD components is also evident, for instance, in the narrowing of the width of resonances A from their free atom results; a forthcoming study of the Fano-shape fitting of all the resonances will reveal the details \cite{javani-future}. On the other hand, significant variations are noted for resonances 1-5 from the coherence. For resonances 1-4, (Ar-$\ful$) and (Kr+$\ful$) exhibit strong asymmetric shapes similar to the resonances in empty $\ful$. But for other two hybrid partners, (Ar+$\ful$) and (Kr-$\ful$), shapes are more nearly symmetric, while the former produces a minimum and the later a maximum. For resonance 5, all four RHA-ICD resonances are asymmetric, although the shape similarity between the bonding hybrid of one system with the anti-bonding of another is retained. Since the excitation channel $nl\rightarrow\eta\lambda$ in \eq{dc-mat-element2} is unchanged, this behavior of shape equivalence between (Ar$\mp$$\ful$) and (Kr$\pm$$\ful$) for 1-5 must depend on the properties of the continuum channel to affect the ICME. Indeed, the primary reason for this behavior lies in the approximate reflection symmetry between the corresponding hybrid wavefunctions. As evident in Fig.\,1(a), multiplying the (Ar-$\ful$) wavefunction by negative 1inverts it to a shape which is close to that of (Kr+$\ful$). One can easily check that this reflection property also holds between the other two hybrids. Obviously, the choice of the caged atom alters the details of the hybridization which subsequently influences the RHA-ICD coherence to determine the resonance shapes. 

\section{Conclusion}
\label{sec4}
In conclusion, we used the TDLDA methodology to calculate a class of innershell-excitation  single-electron autoionizing resonances in the photoionization of Ar@$\ful$ and Kr@$\ful$, decaying into atom-fullerene hybrid final state vacancies. It is demonstrated that these resonances, arising from the interference of the intra-Coulomb autoionizing channel with a coherently admixed inter-Coulomb channel. These resonances are found to be significantly stronger than both regular ICD and Auger resonances, which make them well amenable for experimental detection. The detailed analysis of the results divulge various spectral similarities and differences in the position and shape of the resonances as a function of the central atom. The results indicates that such coherent energy transfer processes must exist across the periodic table when the element supports endofullerene formations, since atom-$\ful$ hybridization is likely to be the rule, not the exception, in the electronic structure of these materials \cite{chakraborty2009xe-hybrid,maser2012zn-hybrid,javani2014cdzn-hybrid}. The current work addresses the {\em participant} decay processes where the excited electron itself drops on to the core-hole. However, the decay of a hole annihilated by a different electron, the {\em spectator} process, can also contribute in a RHA-ICD pathway, suggesting its generality.  Further, these hybrid decay processes are also likely to pervade in the ionization continuum of molecules, nano-dimers and -polymers, and fullerene onion systems that support hybridized electrons as well. In a related context, the attosecond time delay studies of the photoemission of these RHA-ICD resonances can lead to the understanding of the role of electron correlation from a temporal framework which attracted some recent interest \cite{dixit2013timedelay}.   
\begin{acknowledgement}
This work is supported by NSF and DOE, Basic Energy Sciences.
\end{acknowledgement}
%

\begin{thebibliography}{}
%
%
%

\bibitem{cederbaum1997firstTh} 
L.S. Cederbaum, J. Zobeley, and F. Tarantelli, Phys.\ Rev.\ Lett. \textbf{79}, 4778 (1997).

\bibitem{marburger2003firstExp} 
S. Marburger, O. Kugeler, U. Hergenhahn, and T.M\"{o}ller, Phys.\ Rev.\ Lett. \textbf{90}, 203401 (2003).

\bibitem{averbukh2011revTh} 
V. Averbukh, Ph.V. Demekhinb, P. Kolorenc, S. Scheitd, S.D. Stoycheve, A.I. Kuleff, Y.-C. Chiange, K. Gokhberge, 
S. Kopelkee, N. Sisourate, and L.S. Cederbaume, J.\ Electr.\ Spectr.\ Relat.\ Phenom. \textbf{183}, 36 (2011).

\bibitem{hergenhahn2011revExp} 
U. Hergenhahn, J.\ Electr.\ Spectr.\ Relat.\ Phenom. \textbf{184}, 78 (2011).

\bibitem{jahnke2015rev}
T. Jahnke, J.\ Phys.\ B, \textbf{48}. 082001 (2015).

\bibitem{jahnke2004rareDimer} 
T. Jahnke, A. Czasch, M.S. Sch\"{o}ffler, S. Sch\"{o}ssler, A. Knapp, M.K\"{a}sz, J. Titze, C. Wimmer, 
K. Kreidi, R.E. Grisenti, A. Staudte, O. Jagutzki, U. Hergenhahn, H. Schmidt-B\"{o}cking, and R. D\"{o}rner,
Phys.\ Rev.\ Lett. \textbf{93}, 163401 (2004).

\bibitem{oehrwall2004rareClust} 
G. \"{O}hrwall, M. Tchaplyguine, M. Lundwall, R. Feifel, H. Bergersen, T. Rander, A. Lindblad, J. Schulz, S. Peredkov, 
S. Barth, S. Marburger, U. Hergenhahn, S. Svensson, and O. Bj\"{o}rneholm, Phys.\ Rev.\ Lett. \textbf{93}, 173401 (2004).

\bibitem{grieves2011surface} 
G.A. Grieves and T.M. Orlando, Phys.\ Rev.\ Lett. \textbf{107}, 016104 (2011).

\bibitem{jahnke2010water1} 
T. Jahnke, H. Sann, T. Havermeier, K. Kreidi, C. Stuck, M. Meckel, M. Sch\"{o}ffler, N. Neumann, R. Wallauer, S. Voss,
A. Czasch, O. Jagutzki, A. Malakzadeh, F. Afaneh, Th. Weber, H. Schmidt-B\"{o}cking, and R. D\"{o}rner, 
Nat.\ Phys. \textbf{6}, 139 (2010). 

\bibitem{mucke2010water2} 
M. Mucke, M. Braune, S. Barth, M. F\"{o}rstel, T. Lischke, V. Ulrich, T. Arion, U. Becker, A. Bradshaw, and U. Hergenhahn,
Nat.\ Phys. \textbf{6}, 143 (2010).

\bibitem{santra2003triCataion} 
R. Santra and L.S. Cederbaum, Phys.\ Rev.\ Lett. \textbf{90}, 153401 (2003).

\bibitem{kuleff2010multiExcited}
A. I. Kuleff, K. Gokhberg, S. Kopelke, and L. S. Cederbaum, Phys.\ Rev.\ Lett. \textbf{105}, 043004 (2010).

\bibitem{okeeffe2013arDimer}
P. O'Keeffe, E. Ripani, P. Bolognesi, M. Coreno, M. Devetta, C. Callegari, M. Di Fraia, K.C. Prince, R. Richter, M. Alagia,
A. Kivim\"{a}ki, and L. Avaldi, Phys.\ Chem.\ Lett. \textbf{4}, 1797 (2013).

\bibitem{kimura2013arDimer}
M. Kimura, H. Fukuzawa, K. Sakai, S. Mondal, E. Kukk, Y. Kono, S. Nagaoka, Y. Tamenori, N. Saito, and K. Ueda,
Phys.\ Rev.\ A \textbf{87}, 043414 (2013).

\bibitem{fruehling2015time-icd}
U. Fr\"{u}hling, F. Trinter, F. Karimi, J.B. Williams, and T. Jahnke,
J.\ Electr.\ Spectr.\ Relat.\ Phenom. \textbf{204}, 237 (2015).

\bibitem{trinter2013time1} 
F. Trinter, J.B. Williams, M. Weller, M. Waitz, M. Pitzer, J. Voigtsberger, C. Schober, G. Kastirke, C. M\"{u}ller, C. Goihl,
P. Burzynski, F. Wiegandt, T. Bauer, R. Wallauer, H. Sann, A. Kalinin, L.Ph.H. Schmidt, M. Sch\"{o}ffler, N. Sisourat, and T. Jahnke,
Phys.\ Rev.\ Lett. \textbf{111}, 093401 (2013).

\bibitem{schnorr2013time2} 
K. Schnorr, A. Senftleben, M. Kurka, A. Rudenko, L. Foucar, G. Schmid, A. Broska, T. Pfeifer, K. Meyer, D. Anielski, R. Boll, 
D. Rolles, M. K\"{u}bel, M.F. Kling, Y.H. Jiang, S. Mondal, T. Tachibana, K. Ueda, T. Marchenko, M. Simon, G. Brenner, R. Treusch, 
S. Scheit, V. Averbukh, J. Ullrich, C. D. Schr\"{o}ter, and R. Moshammer, Phys.\ Rev.\ Lett. \textbf{111}, 093402 (2013).

\bibitem{gokhberg2014MedApp}
K. Gokhberg, P. Kolorenc, A. I. Kuleff, and L. S. Cederbaum, Nature \textbf{505}, 661 (2014).

\bibitem{trinter2014MedApp} 
F. Trinter, M.S. Sch\"{o}ffler, H.-K. Kim, F.P. Sturm, K. Cole, N. Neumann, A. Vredenborg, J. Williams, I. Bocharova, R. Guillemin, M. Simon,
A. Belkacem, A.L. Landers, Th. Weber, H. Schmidt-B\"{o}cking, R. D\"{o}rner, and T. Jahnke, Nature \textbf{505}, 664 (2014).

\bibitem{gokhberg2006ricdTh} 
K. Gokhberg, V. Averbukh, and L.S. Cederbaum, J.\ Chem.\ Phys. \textbf{124}, 144315 (2006).

\bibitem{barth2005ricdExp1} 
S. Barth, S. Joshi, S. Marbuger, V. Ulrich, A. Lindblad, G. \"{O}hrwall, O. Bj\"{o}meholm, and U. Hergenhahn, 
J.\ Chem.\ Phys. \textbf{122}, 241102 (2005).

\bibitem{aoto2006ricdExp2} 
T. Aoto, K. Ito, Y. Hikosaka, E. Shigemasa, F. Penent, and P. Lablanquie, Phys.\ Rev.\ Lett. \textbf{97}, 243401 (2006).

\bibitem{trinter13HeNe}
F. Trinter, J.B. Williams, M. Weller, M. Waitz, M. Pitzer, J. Voigtsberger, C. Schober, G. Kastirke, C. M\"{u}ller, C. Goihl,
P. Burzynski, F. Wiegandt, R. Wallauer, A. Kalinin, L.Ph.H. Schmidt, M. Sch\"{o}ffler, Y.-C. Chiang, K. Gokhberg, T. Jahnke,
and R. D\"{o}rner, Phys.\ Rev.\ Lett. \textbf{111}, 233004 (2013).

\bibitem{najjari10TwoCenter}
B. Najjari, A.B. Voitkiv, and C. M\"{u}ller, Phys.\ Rev.\ Lett. \textbf{105}, 153002 (2010).

\bibitem{averbukh2006endo-icdTh} 
V. Averbukh and L.S. Cederbaum, Phys.\ Rev.\ Lett. \textbf{96}, 053401 (2006).

\bibitem{korol2011} 
A.V. Korol and A.V. Solov'yov, J.\ Phys.\ B, \textbf{44}, 085001 (2011).

\bibitem{amusia2006} 
M.Ya. Amusia and A.S. Baltenkov, Phys.\ Rev.\ A \textbf{73}, 063206 (2006).

\bibitem{javani2014-rhaicd}
M.H. Javani, J.B. Wise, R. De, M.E. Madjet, S.T. Manson, and H.S. Chakraborty, Phys.\ Rev.\ A \textbf{89}, 063420 (2014).

\bibitem{novotny2012NanoAntenna}
L. Novotny and N. van Hulst, Nature Photonics \textbf{5}, 83 (2012).

\bibitem{madjet2010xeFull} 
M.E. Madjet, T.Renger, D.E. Hopper, M.A. McCune, H.S. Chakraborty, J.M. Rost, and S.~T. Manson,
Phys.\ Rev.\ A \textbf{81}, 013202 (2010).

\bibitem{van1994exchange} 
R. van Leeuwen and E.J. Baerends, Phys.\ Rev.\ A \textbf{49}, 2421 (1994).

\bibitem{ruedel2002oscExp} 
A.R{\"u}del, R. Hentges, U. Becker, H.S. Chakraborty, M.E. Madjet, and J.M. Rost,
\newblock Phys.\ Rev.\ Lett. \textbf{89}, 125503 (2002).

\bibitem{chakraborty2009xe-hybrid} 
H.S. Chakraborty, M.E. Madjet, T. Renger, Jan-M. Rost, and S.T. Manson, Phys.\ Rev.\ A \textbf{79}, 061201(R) (2009).

\bibitem{morscher2010strong}
M. Morscher, A.P. Seitsonen, S. Ito, H. Takagi, N. Dragoe, and T. Greber, Phys.\ Rev.\ A \textbf{82}, 051201 (2010).

\bibitem{madjet-jpb-08} 
M.E. Madjet, H.S. Chakraborty, J.M. Rost, and S.T. Manson, 
\newblock J.\ Phys.\ B \textbf{41}, 105101 (2008).

\bibitem{fano1961}
U. Fano, Phys.\ Rev. \textbf{124}, 1866 (1961).

\bibitem{javani12alkaline-earth} 
M.H. Javani, M.R. McCreary, A.B. Patel, M.E. Madjet, H.S. Chakraborty, and S.T. Manson, Eur.\ Phys.\ J. D \textbf{66}, 189 (2012). 

\bibitem{Seaton83}
M.J. Seaton, Rep.\ Prog.\ Phys.\textbf{46}, 167 (1983) and references therein.

\bibitem{maser2012zn-hybrid} 
J.N. Maser, M.H. Javani, R. De, M.E. Madjet, H.S. Chakraborty, and S.T. Manson, Phys.\ Rev.\ A \textbf{86}, 053201 (2012).

\bibitem{manson69}
S.T. Manson, Phys.\ Rev.\  \textbf{182}, 97 (1969).

\bibitem{kennedy72}
D.J. Kennedy and S.T. Manson, Phys.\ Rev.\ A \textbf{5}, 227 (1972).

\bibitem{theodosiou86}
C.E. Theodosiou, M. Inokuti, and S.T. Manson, At.\ Data Nuc.\ Data Tables \textbf{35}, 473 (1986). 

\bibitem{javani-future}
M.H. Javani,  H.S. Chakraborty, and S.T. Manson, {\em in preparation}.

\bibitem{javani2014cdzn-hybrid}
M.H. Javani, R. De, M.E. Madjet, S.T. Manson, and H.S. Chakraborty,
J.\ Phys.\ B, \textbf{47}, 175102 (2014).

\bibitem{dixit2013timedelay} 
G. Dixit, H.S. Chakraborty, and M.E. Madjet, Phys.\ Rev.\ Lett. \textbf{111}, 203003 (2013).

\end{thebibliography}
%

\end{document}